\documentclass[a4paper,fleqn,usenatbib,english]{mnras}

\usepackage[T1]{fontenc}
\usepackage{ae,aecompl}
\usepackage{amssymb}
\usepackage{amsmath}
\usepackage{babel}
\usepackage{graphicx}
\usepackage{amsmath}
\usepackage{amssymb}
\makeatletter

\newcommand{\sn}[2]{\ensuremath{#1 \times 10^{#2}}}

\newcommand{\rom}[2]{\ensuremath{#1_{\textrm{#2}}}}

\newcommand{\erfc}{\ensuremath{\textrm{erfc}}}
\newcommand{\ebv}{\ensuremath{E(B\,\textrm{--}\,V)}}

\let\oldenumerate\enumerate \renewcommand{\enumerate}{
	\oldenumerate
	\setlength{\itemsep}{1pt}
	\setlength{\parskip}{0pt}
	\setlength{\parsep}{0pt}
}

\let\olditemize\itemize \renewcommand{\itemize}{
	\olditemize
	\setlength{\itemsep}{1pt}
	\setlength{\parskip}{0pt}
	\setlength{\parsep}{0pt}
}

\DeclareMathAlphabet\mathbfcal{OMS}{cmsy}{b}{n}

\title[MCMC+Annealing Photo-z's]{Exploring Photometric Redshifts as an Optimization Problem: An Ensemble MCMC and Simulated Annealing-Driven Template-Fitting Approach}

\author[J. S. Speagle et al.]{
Joshua S. Speagle,$^{1,2,3}$\thanks{E-mail: joshua.speagle@ipmu.jp}
Peter L. Capak,$^{3,4}$
Daniel J. Eisenstein,$^{2}$
Daniel C. Masters,$^{3}$
\newauthor and Charles L. Steinhardt$^{3}$
\\
$^{1}$Kavli IPMU (WPI), UTIAS, The University of Tokyo, Kashiwanoha 5-1-5, Kashiwa, Chiba 277-8583, Japan\\
$^{2}$Harvard University Department of Astronomy, 60 Garden St., MS 46, Cambridge, MA 02138, USA\\
$^{3}$Infrared Processing and Analysis Center, California Institute of Technology, MC 100-22, 770 South Wilson Ave., Pasadena, CA 91125, USA\\
$^{4}$Spitzer Science Center, California Institute of Technology, Pasadena, CA 91125, USA
}

\date{Accepted XXX. Received YYY; in original form ZZZ}

\pubyear{2015}

\begin{document}
\label{firstpage}
\pagerange{\pageref{firstpage}--\pageref{lastpage}}
\maketitle

\begin{abstract}
Using a grid of $\sim$\,2 million elements ($\Delta z = 0.005$) adapted from COSMOS photometric redshift (photo-z) searches, we investigate the general properties of template-based photo-z likelihood surfaces. We find these surfaces are filled with numerous local minima and large degeneracies that generally confound rapid but ``greedy'' optimization schemes, even with additional stochastic sampling methods. In order to robustly and efficiently explore these surfaces, we develop \texttt{BAD-Z} [\textbf{B}risk \textbf{A}nnealing-\textbf{D}riven Redshifts (\textbf{Z})], which combines ensemble Markov Chain Monte Carlo (MCMC) sampling with simulated annealing to sample arbitrarily large, pre-generated grids in approximately constant time. Using a mock catalog of 384,662 objects, we show \texttt{BAD-Z} samples $\sim$\,$40$ times more efficiently compared to a brute-force counterpart while maintaining similar levels of accuracy. Our results represent first steps toward designing template-fitting photo-z approaches limited mainly by memory constraints rather than computation time.
\end{abstract}




\section{Introduction}
\label{sec:intro}


Future large-scale surveys such as \textit{Euclid} \citep{laureijs+11}, the Wide-Field Infrared Space Telescope \citep[\textit{WFIRST}; ][]{green+12}, and the Large Synoptic Survey Telescope \citep[LSST; ][]{ivezic+08} that seek to constrain dark energy equation-of-state using weak gravitational lensing \citep{albrecht+06,bordoloi+12} will require the derivation of redshifts ($z$) to an enormous number of objects ($\gtrsim 10^9$). While spectroscopic redshifts (spec-z's) often are extremely precise, their cost- and time-intensive requirements will necessitate the use of ``photometric redshifts'' (photo-z's) derived from fitting spectral energy distributions (SEDs) taken from a combination of broad- and/or narrow-band photometry in order to measure redshifts to majority of observed objects in a feasible amount of time.

Two main approaches are currently used to derive photo-z's: 
\begin{enumerate}
	\item \textbf{Template fitting}, which attempts to determine the set of \textit{forward} mappings from a collection of model parameters and templates to observed color space.
	\item \textbf{Machine learning}, which attempts to directly determine the best \textit{inverse} mapping from observed color space to redshift via a training set of multi-band photometry and their corresponding spec-z's.
\end{enumerate}

Template fitting-based photo-z codes in use today suffer from several modeling and computational deficiencies. Due to an insufficient understanding of the relevant parameter space, most codes rely on fitting pre-generated ``grids'' of model galaxy photometry to probe corresponding regions of interest. This crude, ``brute-force'' approach not only results in inefficient sampling, but also requires trade-offs in parameter resolution in order to remain computationally viable. As a result, it is generally both too slow \textit{and} too inaccurate to meet the stringent requirements of these future dark energy surveys, even with very sophisticated implementations \citep{brammer+08,ilbert+09}.

Due in part to these issues, many researchers today have turned to machine learning as a way to meet these requirements. While current advances in machine learning-based photo-z's show much promise and perform well with good spectroscopic training sets \citep{carrascokindbrunner13,carrascokindbrunner14,sanchez+14,hoyle15,elliott+15}, the current widespread lack of spectroscopic coverage in specific but relevant regions of color space (Masters et al. 2015, submitted) and calibration issues \citep{cunha+14,newman+15} indicate that template-fitting methods will likely still play a major part in determining good photo-z estimates for these future surveys, especially at higher redshifts where spectroscopic coverage is sparser and more systematically biased.

We attempt to address some of the computational deficiencies involved in template fitting-based photo-z searches. Our main focus is the exploration of the general likelihood surface defined by pre-generated template grids and whether minimization techniques and clever sampling -- both tailored to the general properties of the surface -- can subsequently accelerate photo-z calculation and form the basis of a template-based approach limited by memory availability rather than computation time.

This paper is organized as follows. In \S\ref{sec:formalism}, we give a brief overview of how standard template fitting codes generate model photometry and discuss several approaches for exploring parameter space. In \S\ref{sec:mockcatalog}, we describe the creation of our mock photometric catalog from data in the Cosmological Origins Survey \citep[COSMOS;][]{scoville+07} field that we use for testing, and outline the construction our underlying photo-z model grid. In \S\ref{sec:mapping}, we explore the overall shape of the multidimensional likelihood surface for individual objects. In \S\ref{sec:minimizer}, we use this knowledge to explore the conditional use greedy minimization algorithms, supplemented with stochastic sampling techniques, to quickly locate multidimensional probability modes. In \S\ref{sec:badz}, we present \textbf{B}risk \textbf{A}nnealing \textbf{D}riven Redshifts (\textbf{Z}) (\texttt{BAD-Z}), a Markov Chain Monte Carlo (MCMC)-based approach that combines ensemble MCMC sampling with simulated annealing. In \S\ref{sec:badz_test}, we compare \texttt{BAD-Z}'s performance relative to a corresponding brute-force counterpart across our full mock catalog. Finally, in \S\ref{sec:conclusion} we discuss possible future extensions of this work.

We standardize to the AB magnitude system \citep{okegunn83} throughout the paper.

\section{From Observed SED to Photo-z}
\label{sec:formalism}

Deriving photo-z's from observed photometry is composed of three main parts:
\begin{enumerate} 
	\item Generate model photometry from a list of input parameters $\rom{\vec{F}}{model}(\vec{x})$.
	\item Determine the goodness-of-fit (GOF) between the model photometry and the observed data $\rom{\vec{F}}{obs}$.
	\item Use this information to decide what model parameters to sample.
\end{enumerate}
As understanding each of these components is crucial towards developing more efficient implementations, we discuss each of these in turn.

\subsection{Generating Model Photometry}
\label{subsec:modelphot}

To generate model photometry, most codes begin with a set of ``basis'' galaxy templates which are used either individually \citep{ilbert+09} or in some linear combination scheme \citep{blantonroweis07} to create an underlying galaxy template set. To incorporate the impact of specific emission lines, these galaxy templates are often modified with a set of emission line templates co-added according to a set of scaling relations taken from the literature \citep{ilbert+09,salmon+15}. Rest-frame galaxy templates are created by superimposing an additional \textit{uniform screen} of galactic dust taken from a basis set of normalized dust attenuation curves \citep[see, e.g.,][]{bolzonella+00,charlotfall00} with a given amount of extinction, usually parameterized by $\ebv$.

\begin{figure*}
	\includegraphics[scale=0.385]{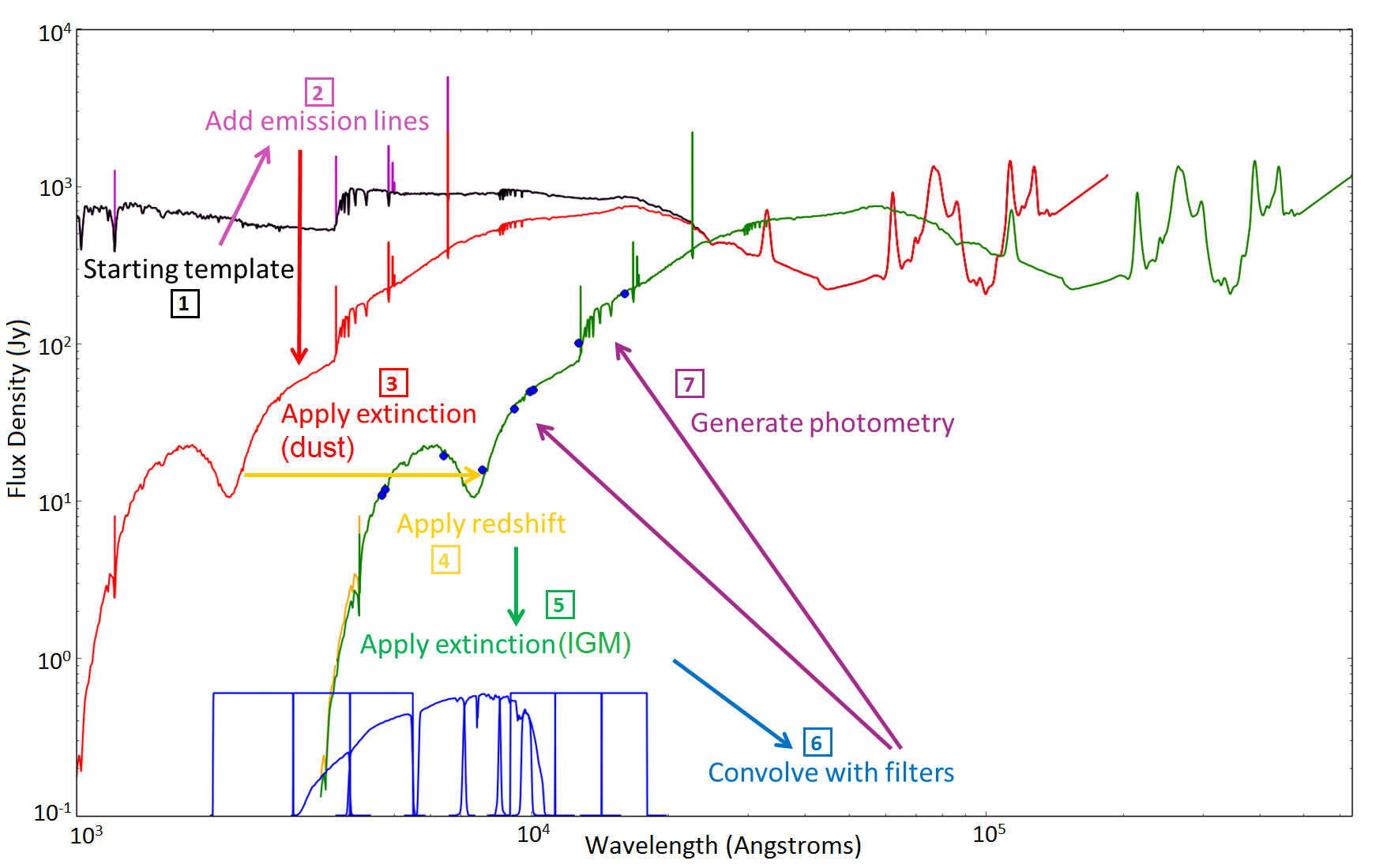}
	\caption{A schematic of how model photometry is generated for a given set of parameters and a collection of galaxy, emission line, and reddening templates. The initial galaxy template (black) is first modified by adding on emission lines (purple) before being reddened by a uniform galactic dust screen (red). The template is then redshifted (orange) and reddening from the IGM is applied (green) before being convolved with a given filter set (blue) to compute the final model photometry (blue circles).}
	\label{fig:gen_phot}
\end{figure*}

These rest-frame galaxy templates are then redshifted by (1+$z$) and modified by extinction from the intergalactic medium (IGM; \citealt{madau95}) to form the final observed-frame galaxy template. The corresponding set of model photometry is then generated by convolving the model galaxy flux density with the transmission of a particular filter (including atmospheric effects) normalized to a source at constant flux density. An illustration of this forward-mapping process is shown in Figure~\ref{fig:gen_phot}.

Generating a new set of model photometry is computationally expensive, involving multiple addition, multiplication, and power/exponential operations on several large ($\gtrsim 10^4$\,--\,$10^5$ element) arrays. Furthermore, although generating a new model template is around an order of magnitude or more computationally taxing than the corresponding filter convolution, the latter process on its own is still too much of a computational burden to be called frequently when computing photo-z's for individual objects.

As a result, often the only computationally feasible approach is to generate a large grid of model photometry for a discrete set of parameters \textit{before} -- rather than during -- the actual SED fitting process. Although other alternatives exist \citep{graff+12,akaret+15}, we focus on optimizing $P(z)$ in the approximately instantaneous likelihood limit (i.e., on a pre-generated grid of model photometry) rather than in the expensive likelihood limit (where the computational overhead involved with choosing samples is much smaller than time necessary to compute the likelihood) due to our emphasis on computational speed.

\subsection{Determining the GOF}
\label{subsec:gof}

In order to derive accurate multidimensional likelihoods, we must choose a suitable GOF metric. Although there are many possible choices, most routines use the simple $\chi^2$ metric in order to incorporate uncertainties on the photometry and/or model in a straightforward manner, where 
\begin{equation}
\chi^2(\vec{x},s) \equiv \sum_i\left(\frac{F_{\textrm{obs},i}-sF_{\textrm{model},i}}{\sigma_i}\right)^2
\end{equation}
where $s$ is an associated model scale-factor, $\sigma_i^2=\sigma_{\textrm{obs},i}^2+\sigma_{\textrm{model},i}^2$ is the total variance, and the sum over $i$ is taken over all observed bands. For a given $\vec{x}$ we can marginalize over $s$ to minimize $\chi^2(\vec{x})$, giving us
\begin{equation}
s = \left. {\sum_i \frac{F_{\textrm{obs},i}F_{\textrm{model},i}}{\sigma_i^2}} \middle/ {\sum_i{\frac{F_{\textrm{model},i}^2}{\sigma_i^2}}} \right.
\end{equation}
This is a simple one-step process that can be calculated prior to computing the actual $\chi^2$ value.

While $\chi^2$ is valid for normally distributed data at any level of signal-to-noise (S/N), it \textit{does not} incorporate any method of treating upper limits, which are relatively common in astronomical data. As a result, many codes (incorrectly) implement ad-hoc procedures in an attempt to include this more limited set of information. This leads to small \textit{but non-negligible} biases in fitting routines for low S/N data involving upper limits.

In this case, a modified $\chi^2$ metric can be derived to properly incorporate upper limits \citep{sawicki12}. For a given set of observations divided into a subset $I$ of observed values $\vec{F}_{\textrm{obs}}$ (included in the sum over $i$) and a subset $J$ of upper limits $\vec{F}_{\textrm{lim}}$ (included in the sum over $j$),
\begin{equation}
\chi^2_{\textrm{mod}}(\vec{x},s) \equiv \chi^2_I(\vec{x},s) + \chi^2_{\textrm{up},J}(\vec{x},s),
\end{equation}
where $\chi^2_I(\vec{x},s)$ is the usual $\chi^2$ statistic, and
\begin{equation}
\chi^2_{\textrm{up},J}(\vec{x},s) \equiv -2 \sum_j \ln \left[\erfc\left( \frac{F_{\textrm{lim},j}-sF_{\textrm{model},j}}{\sqrt{2}\sigma_j} \right) \right]
\end{equation}
is the corresponding term for upper limits, where $\erfc$ is the complementary error function. In this case, there no longer exists a simple analytic expression for $s$ given $\vec{x}$, and the above expression must instead be minimized using numerical methods. As a result, it is \textit{much} more useful to undertake SED fitting in linear flux space, which can accommodate negative fluxes present in low-S/N data, rather than in magnitude space, which cannot.\footnote{``Luptitude'' space \citep{lupton+99} is also a valid alternative.}

\subsection{SED Fitting}
\label{subsec:sedfit}

As outlined in \S\ref{subsec:modelphot}, template fitting methods generate new model photometry by performing a series of operations on a set of templates before convolving the final product with the relevant filter set. As this process is expensive, most current approaches choose to pre-generate a large grid of model templates containing $\sim 10^{6-7}$ individual sets of photometry. Due to the enormous reduction in computing time that can be achieved on a pre-computed grid -- calling $\chi^2$ given $\rom{\vec{F}}{model}(\vec{x})$ is several orders of magnitude faster than generating a new model from scratch -- pre-generating photometry from a large set of parameter combinations often saves orders of magnitude of computation time compared to repeatedly computing them in real time, especially given the large number of objects usually involved.

In order to optimize searches on such grids, we explore three basic classes of methods. The ``brute-force'' approach (i.e., fitting the entire grid) is discussed in \S\ref{subsubsec:grid}, while more adaptive Markov Chain Monte Carlo (MCMC) and nested sampling (NS) methods are discussed in \S\ref{subsubsec:mcmc} and \S\ref{subsubsec:ins}, respectively. We pay particular attention to their specific benefits and drawbacks as well the associated amount of computational overhead.


\subsubsection{Brute-Force Techniques}
\label{subsubsec:grid}

The approach taken by most template-fitting photo-z codes today \citep[e.g., \texttt{Le PHARE};][]{arnouts+99,ilbert+06} is to fit the \textit{entire} pre-generated grid of points to each object to build up a model of the full $N$-dimensional likelihood $P(\vec{x}|\rom{\vec{F}}{obs})$ at a predetermined resolution. Afterwards, maginalized probability distribution functions (PDFs) can be created through 
\begin{equation}
P(\vec{x}_I|\rom{\vec{F}}{obs}) \propto \int P(\vec{x}|\rom{\vec{F}}{obs}) d\vec{x}_J \approx \sum_{\vec{x}_J} P(\vec{x}|\rom{\vec{F}}{obs}),
\end{equation}
where $\vec{x}_I$ are the subset of parameters of interest and $\vec{x}_J$ are the subset of parameters to be marginalized over.


While generally effective \citep{hildebrandt+10, dahlen+13}, brute-force approaches are subject to two major issues:

\begin{enumerate} 
	\item \textit{Inefficient at probing region(s) of interest.} Brute-force methods tend to spend the majority of time for any given object ($\gtrsim\,99\%$) sampling regions of extremely low probability.
	\item \textit{Scale proportional to dimensionality of problem.} The corresponding size of the grid increases multiplicatively with the number of dimensions to be probed and the desired granularity of each dimension.
\end{enumerate}

%

These caveats aside, however, implementations of brute-force approaches can often run relatively quickly compared to other methods that might nominally sample more efficiently due to the straightforward setup and lack of any real computational overhead, which can take full advantage of parallel processing as well as vectorization of repeat operations.

\subsubsection{Monte Carlo Markov Chain-based Techniques}
\label{subsubsec:mcmc}

Unlike grid-based approaches, MCMC-based algorithms \citep[see, e.g., \texttt{SatMC};][]{johnson+13} sample at a rate proportional to the PDF itself and weight every \textit{accepted} trial point evenly \citep[although see ][]{bernton+15}. A standard search heuristic employed by most MCMC codes is the Metropolis-Hastings algorithm \citep{metropolis+53,hastings70}:
\begin{enumerate} 
	\item Draw a set of trial parameters $\vec{x}_n$ from the neighborhood function\footnote{Also often referred to as a ``proposal distribution''.} $q(\vec{x}|\vec{x}_{n-1})$.
	\item Accept the new trial and move to location $\vec{x}_n$ with probability $\min \left( 1,\frac{P(\vec{x}_n|\vec{F})}{P(\vec{x}_{n-1}|\vec{F})}\frac{q(\vec{x}_n|\vec{x}_{n-1})}{q(\vec{x}_{n-1}|\vec{x}_n)} \right)$. Otherwise, remain at $\vec{x}_{n-1}$.
	\item Repeat from step (i) until a stopping criterion is reached.
\end{enumerate}
This procedure is used to guide several individual ``chains'' of related draws as they converge (``burn in'') to and eventually begin sampling from the region of interest. Although the exact neighborhood function chosen is ultimately arbitrary, most often an $N$-dimensional multivariate-normal distribution $N(\vec{\mu},\mathbf{\Sigma})$ is used, where $\vec{\mu}=\vec{x}_{n-1}$ is the mean vector (adjusted at each step) and $\mathbf{\Sigma}=\vec{\sigma}^2\mathbf{I}$ is the covariance matrix ($\mathbf{I}$ being the identity matrix), also often adjusted over the course of a typical run. 

Because MCMC-based algorithms sample at a rate approximately proportional to the PDF in a given region of interest, they are able to explore large, $N$-dimensional spaces at finer resolution and with far fewer function calls than grid-based approaches while scaling relatively slowly with the dimensionality of the problem \citep[up to moderate dimensionality;][]{handley+15}. However, MCMC-based approaches are still subject to several issues:
\begin{enumerate}
	\item \textit{Dependent on the form of the neighborhood function.} The performance of MCMC-based algorithms is sensitive to the precise construction and implementation of the neighborhood function.
	\item \textit{Only samples locally.} MCMC approaches cannot reliably recover multimodal distributions where the modes are separated by distances much larger than the neighborhood function.
	\item \textit{Fails to generate fully independent draws.} Due to their random-walk nature, individual chains should be ``thinned'' to ensure fully independent random samples.
	\item \textit{Ineffective during burn-in.} Trials taken during the burn-in phase cannot be used to derive the PDF because they are sensitive to the chains' initial positions.
\end{enumerate}

An MCMC-based approach running on a large, pre-generated grid represents a more efficient way of developing a model of the likelihood by exploring only the portion of the grid that contains the majority of the probability \citep[see, e.g., \texttt{SpeedyMC};][]{acquaviva+12}. As MCMC algorithms only involve a small amount of computational overhead\footnote{Assuming the neighborhood function remains fixed. If it is instead allowed to adapt during the burn-in phase, the amount of computational overhead becomes somewhat larger.} for each likelihood evaluation, large efficiency gains can be achieved while retaining similar levels of accuracy to brute force approaches. In addition, since MCMC methods generally tend to sample in approximately constant time at fixed dimensionality, arbitrarily large grids can be used without substantially increasing the computation time, provided they can be loaded into memory.

\subsubsection{(Importance) Nested Sampling-based Techniques}
\label{subsubsec:ins}

Importance Nested Sampling \citep{ferozhobson08,feroz+09,feroz+13} provides an interesting approach to deal with the degeneracies observed in photo-z parameter space, as the method is designed to sample multimodal likelihoods. During INS, $\rom{N}{live}$ ``live'' points are drawn from the prior $P(\vec{x})$. At each subsequent iteration $i$, the point with the lowest $\mathcal{L}_i$ is removed from the set and replaced with another drawn from the prior under the constraint $\mathcal{L}_{i+1} > \mathcal{L}_i$. In order to efficiently draw unbiased samples from the likelihood-constrained prior, at each iteration $i$ the full set of $\rom{N}{live}$ points is decomposed to lie within a set of (possibly overlapping) ellipsoids that can be evolved separately, allowing the algorithm to probe widely separated minima. As this sampling process is repeated, the live particles move through ``nested'' shells of constrained likelihood as the prior volume contained in the union of ellipsoids is steadily reduced, finally terminating after reaching a specific tolerance threshold. An estimate of the posterior can then be obtained from the entire set of trials via using a corresponding set of weights  \citep{cameronpettitt13}, as outlined in \citet{feroz+13}. 



\begin{figure*}
	\includegraphics[scale=0.40]{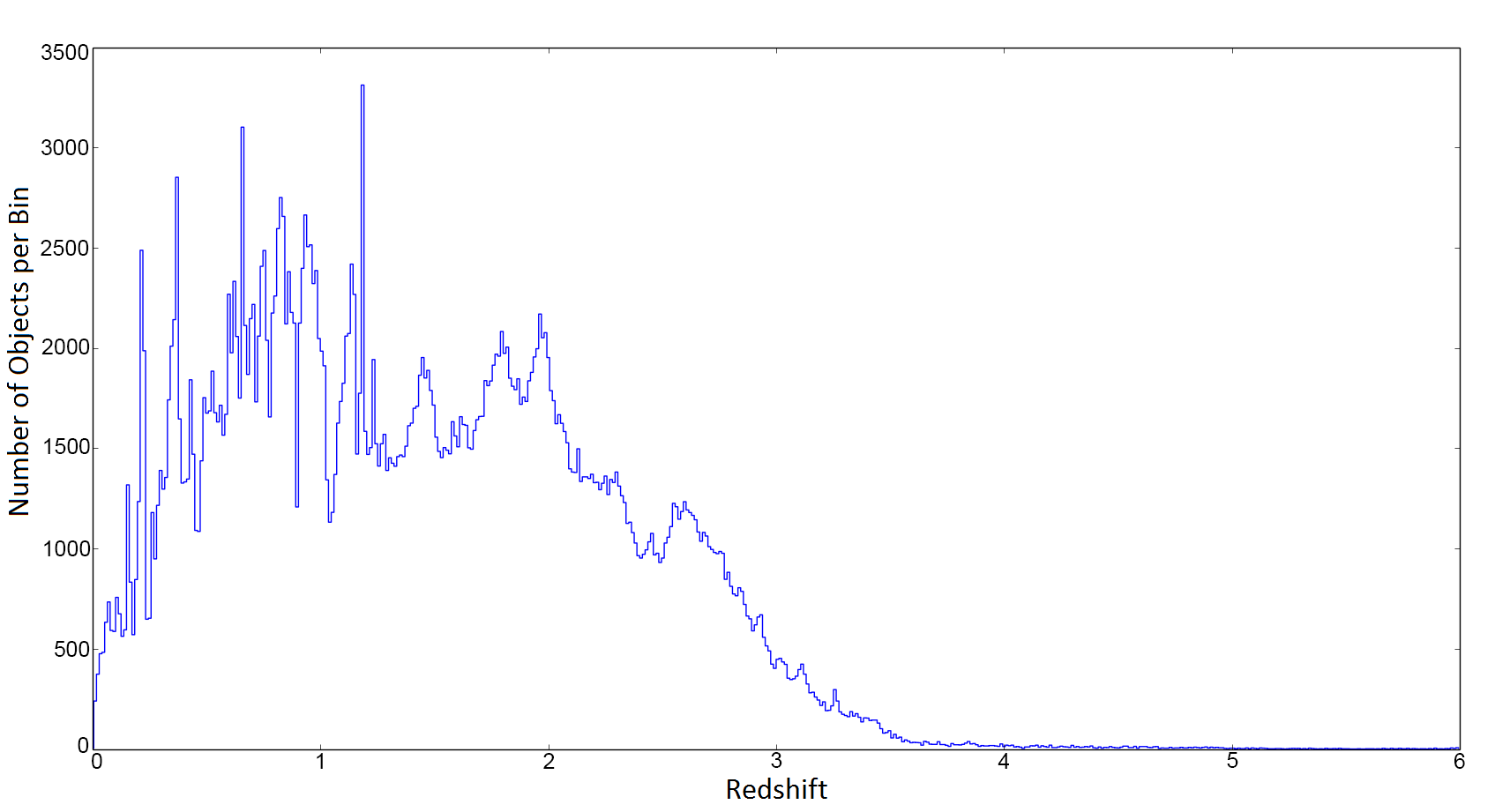}
	\caption{The redshift distribution of our mock catalog of 384,662 COSMOS galaxies. Although the distribution extends out to $z=6$, the majority of galaxies are located at $z \lesssim 3.2$.}
	\label{fig:redshift_dist}
\end{figure*}

\begin{figure*}
	\includegraphics[scale=0.41]{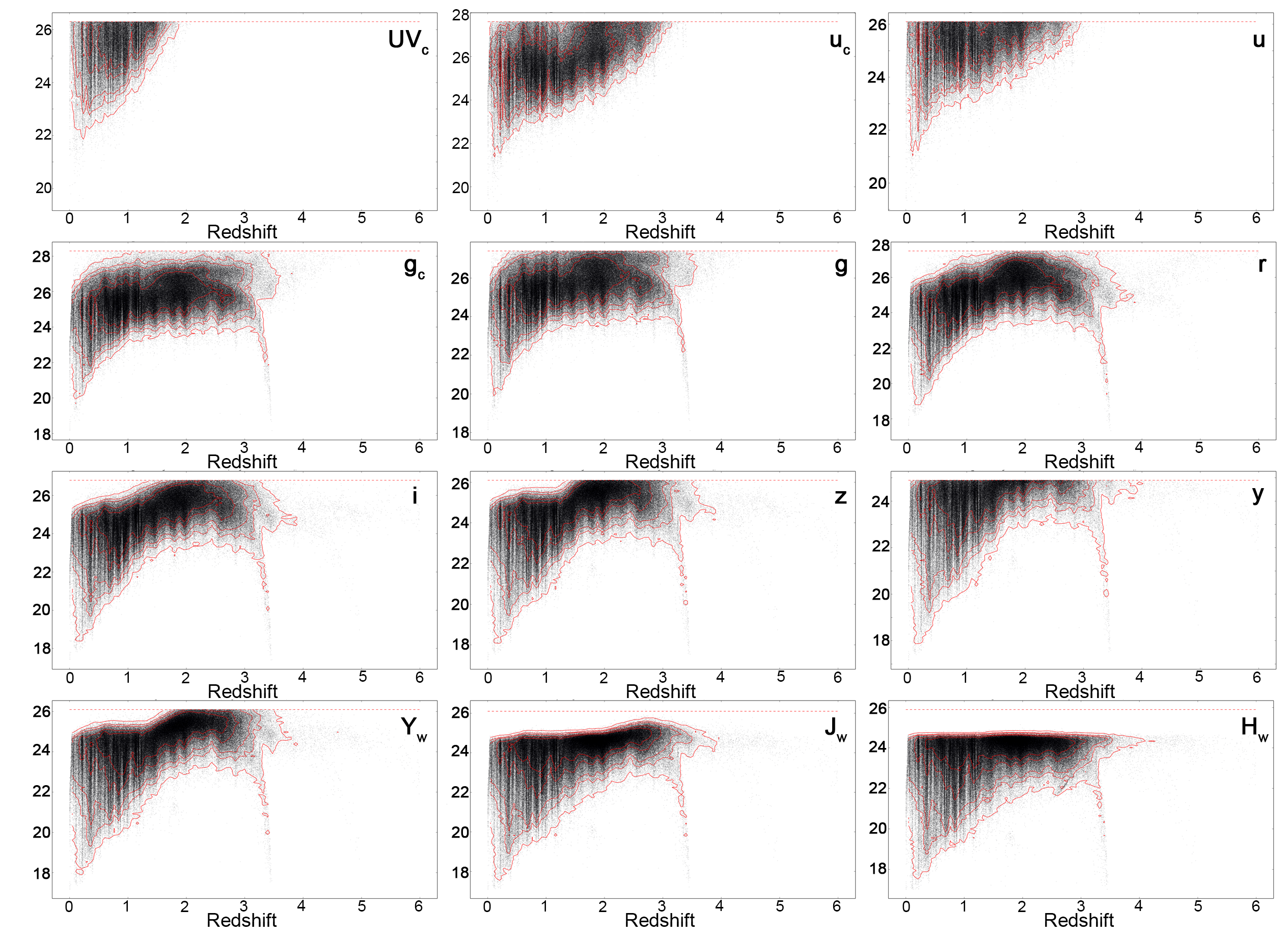}
	\caption{Photometry (in magnitudes) in each band for all objects in our mock COSMOS catalog as a function of redshift, with 50\%, 70\%, 90\%, 95\%, and 99\% contours overplotted in red. 5-$\sigma$ upper limits are denoted by horizontal dashed red lines. Our mock catalog spans a wide range of magnitudes and imaging depths in all available bands. Our $H=24.5$\,mag cut (to mimic the imaging depth of the \texttt{Euclid} wide-field survey) can be seen in the bottom-right panel.}
	\label{fig:combined_phot_dist}
\end{figure*}

While INS enables the determination of more accurate PDFs by capturing degeneracies that might otherwise be missed by MCMC-based approaches, it is substantially slower due to the large computational overhead involved with computing the ellipsoidal decompositions (large) and the associated weights (moderate) for a pre-defined number of iterations. As a result, we limit our investigation in \S\ref{sec:badz} to MCMC-based algorithms, and defer analysis of INS-based approaches to possible future work.

\section{Generating a Realistic Mock Catalog}
\label{sec:mockcatalog}

To avoid possible issues caused by template mismatch (especially when compared to the full range of observed galaxy SEDs, including emission line variation) and other possible systematic effects, we opt to explore the parameter space spanned by a mock catalog of galaxies using the same set of templates we hope to later test.

We begin the construction of our mock catalog using the high-quality photometry ($\sim$\,30 bands) available in the COSMOS field \citep{capak+07}. Photo-z's are derived with the brute-force template-fitting code \texttt{Le PHARE} using a grid with the following specifications:
\begin{enumerate}
	\item $N_z = 601$: Spans the redshift range $z=0$\,--\,$6$ in steps of $\Delta z = 0.01$.
	\item $\rom{N}{T,gal} = 31$: Includes 8 elliptical templates and 11 spiral templates (constructed by linearly interpolating between the \citet{polletta+07} templates), supplemented with 12 starburst (SB) templates constructed from \citet{bruzualcharlot03} SPS models assuming exponentially declining SFHs with ages ranging from $3$ to $0.03$\,Gyr.
	\item $\rom{N}{emline} = 3$: Only a single template including Ly$\alpha$, O{\scriptsize[II]}, H$\beta$, O{\scriptsize[III]}, and H$\alpha$ is used. This is added to each template prior to applying reddening effects according to $\lbrace 0.5$, $1.0$, $2.0\rbrace$ times the scaling relations outlined in \citet{ilbert+09}.
	\item $\rom{N}{T,dust} = 5$: The dust curves used include observations from the SMC \citep{prevot+84}, LMC \citep{fitzpatrick86}, MW \citep{seaton79,allen76}, and SB galaxies \citep{calzetti+00}.
	\item $N_{\ebv}=9$: Each template is allowed $\ebv$ values of $\lbrace 0$, $0.05$, $0.1$, $0.15$, $0.2$, $0.25$, $0.3$, $0.4$, $0.5\rbrace$.
\end{enumerate}
After correcting for zero-point offsets using the spectroscopic sample from zCOSMOS \citep{lilly+07,lilly+09}, a subset of the grid is then fit to each object and marginalized over to derive the associated redshift PDF. See \citet{ilbert+09} for additional information.


Using the median $P(z)$ value (regenerated to higher $\Delta z$ resolution than the underlying grid), best-fitting template, reddening law, $\ebv$, and scaling factor derived from \texttt{Le PHARE} for each galaxy, we create a collection of corresponding model galaxy templates that mimic the observational data.\footnote{Due to the coarse nature of the emission line grid, we do not include emission line contributions to simplify the nature of our tests. This does not impact our conclusions, although we explore ways of incorporating this additional complexity in future work (Speagle et al. 2015, in preparation).}
We convolve these templates with a set of 12 filters designed to mimic the wavelength ranges probed by the future Cosmological Advanced Survey Telescope for Optical and UV Research (\textit{CASTOR}; UV$_c u_c g_c$) and \textit{Euclid} ($Y_wJ_wH_w$) missions supplemented with ground-based photometry ($ugrizY$). These provide a wide wavelength range to detect spectral features, including 3 overlapping bands ($u$, $g$, and $Y$).

From an initial sample of $\sim$\,2 million galaxies, we implement an $H_w=24.5$\,mag cut to mimic the 5-$\sigma$ imaging depth of the wide-field \textit{Euclid} survey. This leaves us with $\sim$\,$20\%$ of the original sample, or 384,662 galaxies spanning a redshift range from $z=0$ to $\sim$\,$3.2$, including a small number of objects up to $z \sim 6$ (Figure~\ref{fig:redshift_dist}).

These photometric fluxes are then jittered according to the expected background noise levels based on the anticipated depth of the imaging in each band\footnote{This calculation does not include error from shot noise from galaxy photons, which is expected to be on the order of $\sim$\,$2\%$ at the $3\sigma$ detection limit.} to create the final mock catalog. The distribution of photometry (in magnitudes) as a function of redshift in each of the filters is shown in Figure~\ref{fig:combined_phot_dist} along with their respective 5-$\sigma$ imaging depths. To avoid complications arising from the inclusion of upper limits (see \S\ref{subsec:gof}), we leave the model photometry in flux space.

In addition, we also compute a baseline GOF value for each object ($\rom{\chi^2}{base}$) by determining the associated $\chi^2$ values between the original (pre-jitter) model photometry and final (post-jitter) mock photometry. This allows us to see the extent to which our errors altered the original GOF while also providing a check on the overall quality of the best fits determined by our code(s). In particular, given optimal performance and an ideal model set, the derived $\chi^2_{\textrm{obj}}$ for an object should always satisfy the condition $\chi^2_{\textrm{obj}} \leq \chi^2_{\textrm{base}}$. 

In other words, an ideal photo-z code should always find either (1) the ``correct'' solution or (2) an ``incorrect'' one that is a better fit to the data. This is not always true in practice, especially in cases (such as ours) where the associated redshift has been regenerated to a higher resolution than the corresponding $\Delta z$ spacing of the grid. While being able to find the ``best'' match to the data is a desirable feature of an effective fitting routine, it is not strictly necessary -- as long as the general region \textit{around} the best fit is probed sufficiently well (i.e., the relative \textit{shape} of the PDF can be recovered even when sparsely sampled), a given algorithm should still be able to derive accurate $P(z)$'s and associated estimates.

\begin{figure*}
	\includegraphics[scale=0.26]{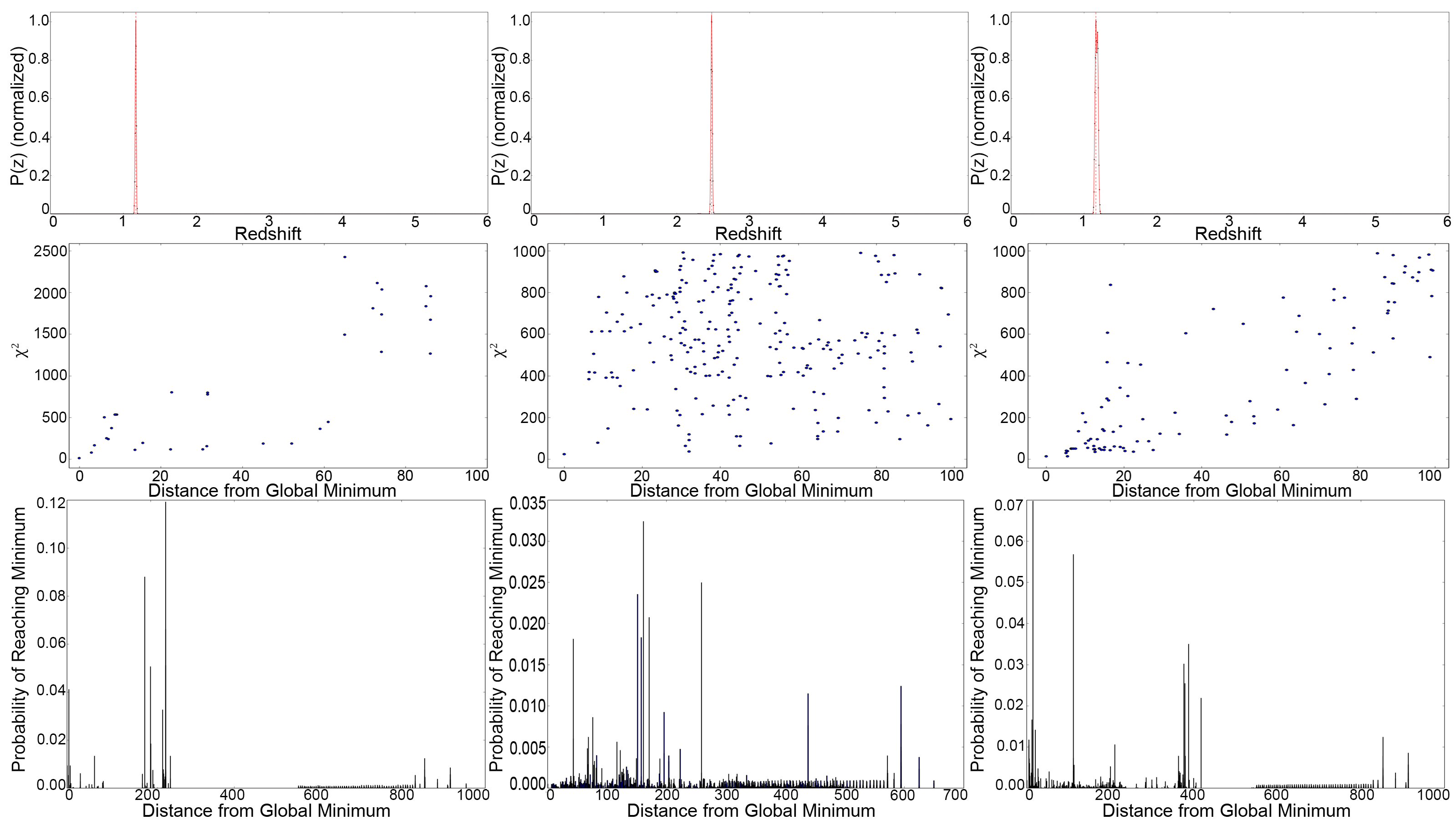}
	\caption{4-D maps of the photo-z likelihood surface for three representative objects from our mock catalog shown from left to right.
		\textbf{Top:} Marginalized $P(z)$ distributions for each object. Individual grid points are plotted as black dots, with a cubic spline fit shown with a solid red line and the location of the peak marked by a dashed red line. In all cases $P(z)$ distribution is narrow and well-defined.
		\textbf{Middle:} The $\chi^2$ values of individual minima plotted as a function of the Euclidean distance from the \textit{global minimum}. Each plot shows all minima found within $\Delta z \sim 0.5$ ($\Delta z =0.1$ is $\sim$\,20 grid units). The large number of minima indicates the region around the global minimum value is quite bumpy. However, since most minima have significantly higher $\chi^2$ values than the best-fit, they tend to have an extremely small impact on the marginalized $P(z)$'s (top panels). Deriving accurate $P(z)$'s thus relies heavily on being able to locate and sample the small region surrounding the global minimum.
		\textbf{Bottom:} The probability (normalized to 1) that a random trial point on the grid will reach a specific minimum, plotted as a function of Euclidean distance from the global minimum. While the global minimum is a far better fit than most competing minima (middle panels), the latter occupy large areas of parameter space, making it extremely difficult for local and/or gradient-based minimization algorithms to reach relevant $P(z)$ modes. The impact of degeneracies at widely separated distances from the global minimum (clusters of large minima at $\rom{d}{grid} \gtrsim 100$) as well as edge effects (``furrows'' at $\rom{d}{grid} \gtrsim 300$) can also be seen, both of which further impede this process.}
	\label{fig:4d_map}
\end{figure*}

\section{Mapping the Photo-z Likelihood Surface}
\label{sec:mapping}

Using a finely spaced grid of $\sim$\,2 million sets of model photometry -- identical to the one used by \texttt{Le PHARE} except with an improved $\Delta z$ resolution of $0.005$ -- we calculate the $\chi^2$ value at every trial point for a series of individual objects from our mock catalog. We then derive the corresponding locations and likelihoods for competing minima as a function of the Euclidean distance (normalized to the grid spacing in each dimension) from the global minimum. To determine the relative ``size'' of each minimum, we force each point on the grid follow the surrounding (discrete) gradient until it locates the closest corresponding minimum and record the final number of trial points occupying each local minimum.

To understand how this underlying structure corresponds to the output $P(z)$ distribution, we marginalize over all trials to derive $P(z)$. Together, these pieces of information not only inform us about the approximate general distribution, size, depth, and behavior of minima within the 4-D parameter space, but also how the corresponding structure affects the final marginalized distribution of interest. The results for three representative objects are plotted in Figure~\ref{fig:4d_map}.

We find several main features of interest:
\begin{enumerate}
	\item The main global minimum is surrounded by numerous competing minima that occupy sizeable regions of parameter space (middle panels), indicating that the space directly in the area of the global minimum and other notable degeneracies (e.g. confusion over the $1216$\,{\AA} and $4000$\,{\AA} breaks) are quite ``bumpy''.
	\item The redshift-reddening degeneracy (i.e., red objects may either be dusty or at high(er) redshift) appears to occupy a significant region of parameter space (bottom panels), although it's overall contribution to the marginalized likelihood is (usually) small (top panels).
	\item A series of fitting artifacts (i.e., edge effects) due to highly mismatched templates lead to small ``ridges'' within the corresponding region of parameter space (bottom panels) that have negligible probability but occupy a small  region of parameter space that needs to be avoided.
\end{enumerate}
Most crucially, almost none of this substructure appears in the final redshift PDF (top panels). As a result, not only does the full 4-D space appear to contain hundreds to thousands of local minima with degenerate regions a significant distance away from the global minimum occupying large regions of parameter space, but most of this structure remains hidden when investigating marginalized 1-D $P(z)$ distributions only.

Further investigations in Speagle et al. (2015, in preparation) suggest that much of this substructure is due to the combination of 1-D projections of non-linear effects in the creation of the grid. Galaxy SEDs occupy a very non-linear $N$-dimensional manifold in color space, with complex and correlated changes in the overall shape. However, when used in most grids, they are often assigned arbitrary numbers based on some simple diagnostics (e.g., FUV flux), equivalent to a crude 1-D projection. This projection loses valuable information concerning the higher-dimensional manifold, and when combined with the equivalently crude projection of dust attenuation curves (where the ordering is often equally arbitrary) and redshift evolution, creates a number of (wide) local minima that might otherwise not exist. While this indicates that a more suitable choice of grid might eliminate some of the bumpiness and/or edge effects, the broad features observed in Figure~\ref{fig:4d_map} still remain valid in general.


\begin{figure*}
	\includegraphics[scale=0.68]{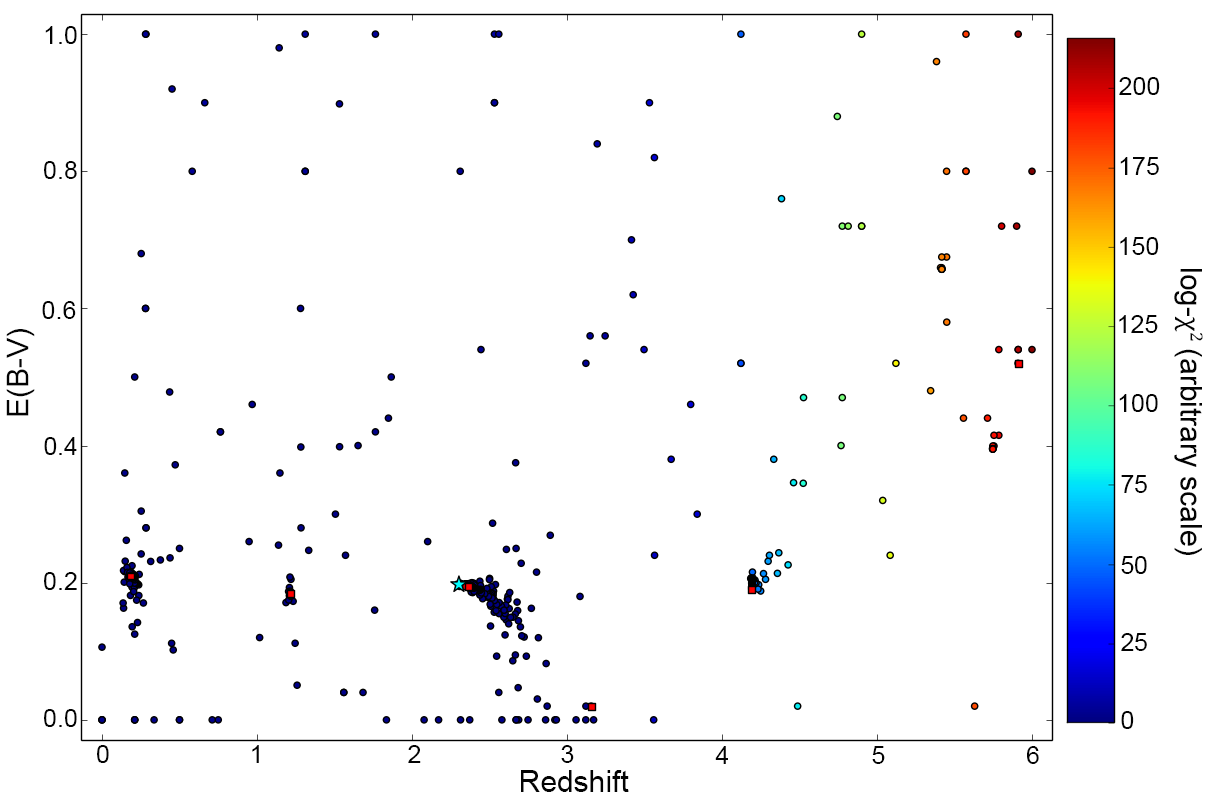}
	\caption{COBYLA minimization runs using stochastic basin-hopping ($\rom{N}{hops}=10$) on a 2-D $z$--$\ebv$ slice for a sample object at $\left[z,\ebv\right]=\left[2.2221,0.2\right]$ given the ``correct'' underlying galaxy-dust template combination. Each trial point is plotted and colored according to the corresponding scaled $\log$-$\chi^2$ value while the best value located during each run is marked using red squares. As Figure~\ref{fig:4d_map}, we find the reduced subspace retains a significant number of local minima surrounding the global best-fit value (cyan star), a significant confounding factor for standalone uses of ``greedy'' minimization algorithms on pre-generated grids of model photometry.}
	\label{fig:2d_map}
\end{figure*}

\section{Optimization in Redshift-Reddening Space}
\label{sec:minimizer}

Although the general likelihood surface is relatively rough, smaller dimensional slices might be smoother and thus more amenable to ``greedy'' (i.e., strict gradient-descent) optimization methods \citep[e.g., Levenberg-Marquardt;][]{levenberg44,marquardt63} that could be used to accelerate photo-z calculations. Since many of the most effective greedy algorithms used in higher-dimensional minimization problems only function properly for a (quasi-)continuous set of input parameters, to avoid the difficulty in interpolating between discrete templates we explore their conditional use on reduced 2-D $z$--$\ebv$ subspaces for a given galaxy template and dust attenuation curve.

Since we are minimizing quantities that are both non-negative and bounded, we focus on the use of constrained minimization algorithms. Although the results presented in this section use Constrained Optimization by Linear Approximation \citep[COBYLA;][]{powell94}, we test that they remain largely unchanged if other (un)constrained minimization algorithms (and associated penalty functions) are used.

In order to investigate the possibility of multiple, widely-separated minima within the subspace, we supplement our chosen greedy minimizer with a simple version of ``basin-hopping'' \citep{walesdoye97}, an iterative stochastic \textit{metaheuristic} (i.e., a heuristic that uses pre-existing search heuristic) designed to probe spaces with a few deep but widely separated degeneracies. It acts on a given minimization routine as follows:
\begin{enumerate}
	\item Given the position $\vec{x}_{\textrm{min},i}$ of the last accepted minimum, randomly jump to a new coordinate $\vec{x}_{\textrm{start},j}$ based on a given neighborhood function $q(\vec{x}|\vec{x}_{\textrm{min},i})$.\footnote{As with MCMC approaches, this is often chosen to be an $N$-dimensional multivariate Gaussian.}
	\item From $\vec{x}_{\textrm{start},j}$, use a chosen minimization algorithm to find the nearest minimum $\vec{x}_{\textrm{min},j}$.
	\item Replace $\vec{x}_{\textrm{min},i \rightarrow i+1}$ with the new coordinates $\vec{x}_{\textrm{min},j}$ according to the Metropolis-Hastings criterion.
	\item Repeat from step (i) until a stopping criterion is reached.
	\item Select the global minimum from the collection of all accepted minima $\lbrace\vec{x}_{\textrm{min},i}\rbrace$.
\end{enumerate}
This procedure is conceptually similar to an MCMC algorithm, where instead of calling a likelihood function at each iteration we run a specific minimization routine starting from that location. To avoid any procedural fine-tuning, we let $q(\vec{x})$ be uniform over the entire $z$ and $\ebv$ interval and only use a fixed number of $\rom{N}{hops}$ runs.

The output for one such object given the ``correct'' (i.e., intrinsic) galaxy and dust templates is shown in Figure~\ref{fig:2d_map}. We find the total number of trials required to find the global minimum in this subspace is relatively small, with $\rom{N}{trials} \lesssim 1500$ for $\rom{N}{hops}=10$. This corresponds to $\sim 30\%$ of the size of the corresponding coarse $z$--$\ebv$ grid, albeit sampled at arbitrary (rather than fixed) resolution. As $\rom{N}{trials}$ scales approximately linearly with $\rom{N}{hops}$, the true efficiency for an algorithm tailored to the specifics of this space would likely be somewhat higher. As a result, we find that running a constrained stochastic minimization procedure for every combination of discrete parameters by itself is at least a factor of three more efficient at locating the global minimum (with improved accuracy) than a standard grid search.

Unfortunately, we find that many of the unwanted features present in our full 4-D map (Figure~\ref{fig:4d_map}) are also present in this reduced subspace.\footnote{These features are also not limited to a specific galaxy-dust template combination -- for a similar procedure computed using a slightly mismatched galaxy template-dust curve combination, we observe the same general features. For extremely mismatched templates (e.g., trying to fit an elliptical to a slightly reddened spiral), however, we instead find the algorithm repeatedly terminates at boundary locations (i.e., grid edges) such as $\left[z,\ebv\right] = \left[0,0\right]$.} As a result, we conclude that the number of basin-hopping trials necessary to ensure that greedy algorithms will reliably converge to the global minimum is likely quite large, negating most of the potential gains it may have provided.

\section{Designing an Efficient and Robust Photo-z Search Algorithm}
\label{sec:badz}

Based on these results and the discussion in \S\ref{subsec:sedfit}, we now explore the use of MCMC sampling to explore large pre-generated grids quickly and efficiently. We find that the ensemble MCMC sampling implementation from \citet{goodmanweare10} and implemented in \texttt{emcee}\footnote{\hyperref[http://dan.iel.fm/emcee]{http://dan.iel.fm/emcee}} \citep{foremanmackey+13} addresses most of the issues regarding MCMC approaches outlined in \S\ref{subsubsec:mcmc} while avoiding many of the drawbacks. As a result, we opt to use it as the underlying foundation of our algorithm and describe the basic implementation below.

\begin{figure*}
	\includegraphics[scale=0.25]{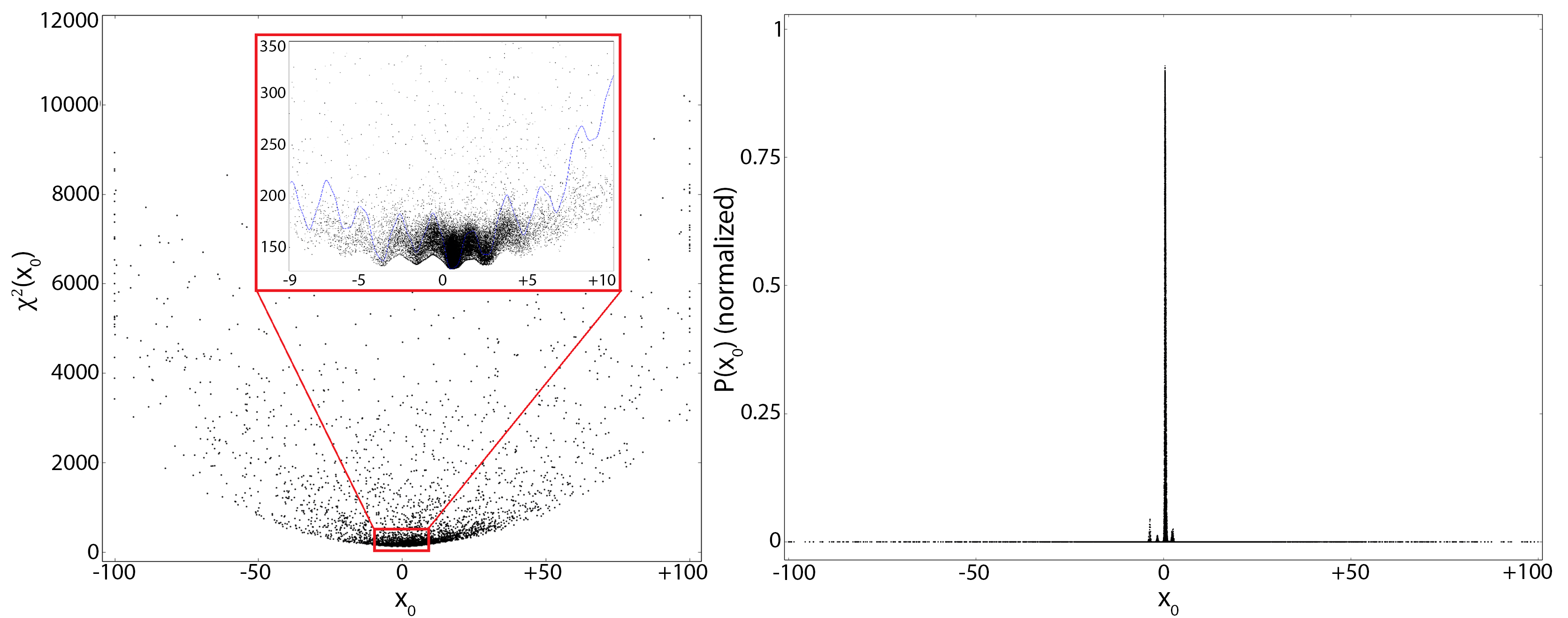}
	\caption{Individual trials as a function of position in terms of $\chi^2$ (left) and $P=e^{-\frac{\chi^2}{2}}$ (right) for a 1-D projection of a toy 4-D function designed to broadly mimic the features of our photo-z space described in \S\ref{sec:mapping} and observed in Figure~\ref{fig:4d_map}. A portion of a 1-D component of this function is shown in blue (left central inset). The dense sampling in regions surrounding the global minimum illustrates that the algorithm is sampling effectively and can robustly locate regions of high probability in significantly bumpy parameter spaces.}
	\label{fig:sample_function_sampling}
\end{figure*}

\begin{figure*}
	\includegraphics[scale=0.3]{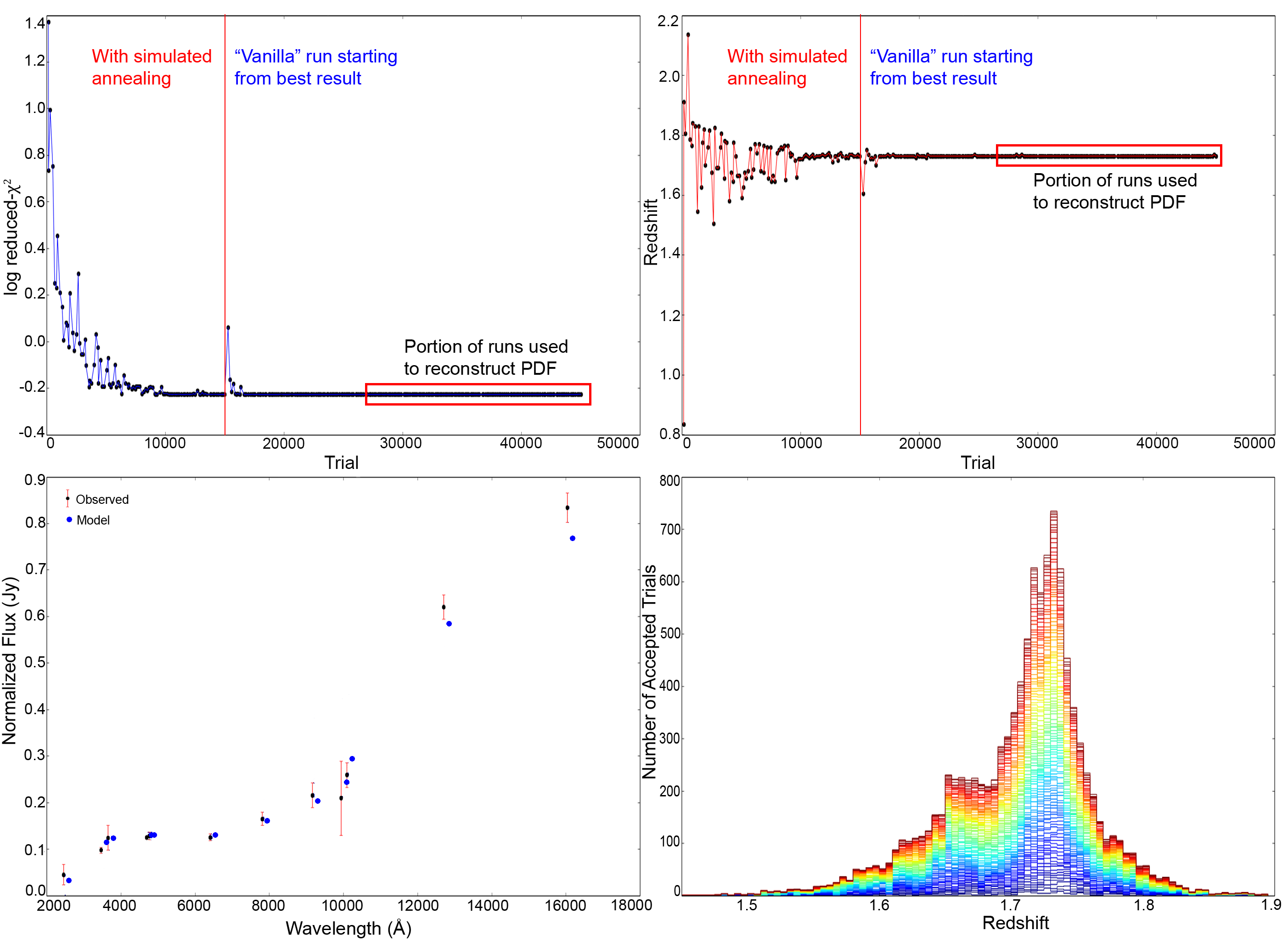}
	\caption{
		\textbf{Top left:} The minimum $\log (\textrm{reduced-}\chi^2$ found among the full set of walkers after each iteration through the entire ensemble for an random object from our mock catalog. In this specific case, \texttt{BAD-Z} converges to the best-fit set of parameters $\rom{\vec{x}}{best}$ after $\sim 60\%$ of the simulated-annealing driven portion of the run ($0.6\%$ the size of the corresponding grid), and relocates it relatively quickly after the walkers have re-spawned for the ``vanilla'' ($T(t)=1$) portion. 
		\textbf{Top right:} As the top left panel, but for the best-fit redshift $\rom{z}{best}$. Similar behavior can be seen, although small perturbations away from $\rom{z}{best}$ are now visible. 
		\textbf{Bottom left:} The best-fit model (blue circles) versus observed (black circles with red error bars) photometry plotted at effective wavelength in each filter. For clarity, the model photometry has been slightly shifted. The observed agreement between the two sets of points indicates the overall quality of the final fit(s) as seen in the top left panel. 
		\textbf{Bottom right:} The redshift PDF $P(z)$ derived from the final $40\%$ of all trials as a function of time (blue to red). As we are consistently probing the area directly around $\rom{\vec{x}}{best}$ (see top left), \texttt{BAD-Z} is able to effectively reconstruct the underlying PDF using only 18,000 independent trials, or about $1.1\%$ of the size of the corresponding grid.
	}
	\label{fig:badz_run}
\end{figure*}


In brief, rather than spawning $M$ chains that sample from a neighborhood function, we instead spawn a (much larger) ensemble of $N \gg M$ ``walkers'' which each sample from on the current locations of the complementary ensemble (i.e., the other $N-1$ walkers) rather than any individual walker's local neighborhood. By eliminating the neighborhood function in this manner, we decrease the fine-tuning often required for MCMC approaches while also ensuring that every trial is nearly independent, simultaneously solving a number of issues raised in \S\ref{subsubsec:mcmc}. For our purposes then, ensemble MCMC sampling gives all the benefits inherent in an MCMC-driven approach on a pre-computed model grid while eliminating the majority of the drawbacks.

However, although the general implementation outlined above is powerful for sampling around the region of interest, it will still be relatively inefficient during the burn-in phase. To assist with this, we turn to \textit{simulated annealing}, a metaheuristic designed to assist searches such as these where the goal is to find the global minimum in a bumpy and often significantly multimodal space. The overall method involves imposing a global temperature on either the entire space (the standard implementation) or individual samplers/regions \citep[otherwise known as ``tempering''; see][]{johnson+13} that distorts the shape of the space such that
\begin{equation}
P(\vec{x})\rightarrow \left[P(\vec{x})\right]^{T_0/T(t)},
\end{equation}
where $T(t)$ is the temperature as a function of time (i.e. iteration) and $T_0$ is the ``transition temperature'' that we henceforth take to be $1$.

For $T(t)>T_0$, ``bad'' jumps have a higher relative probability of being accepted, allowing an algorithm additional stochasticity while sampling. As a result, it is able to explore more of the search space in the hopes of finding the region around the global minimum. For $T(t)<T_0$, bad jumps instead become relatively \textit{less} likely of being accepted, reducing stochasticity and increasingly constraining the algorithm to only move in the direction of the gradient. Ideally, this forces our algorithm to find a more optimal solution assuming that it has reached the general region of interest.

To test whether that this approach will improve burn-in performance in the region we are hoping to explore, we investigate the performance of a simulated annealing-driven MCMC-based algorithm on a sample function designed to mimic the overall properties of the photo-z parameter space we observed in \S\ref{sec:mapping}. Using an ensemble of 250 walkers, $a=2$, $T(t=0)=2.5$, and $\Delta T=0.01$ per ensemble run (i.e. after iterating over all walkers once), we verify that not only is the combined algorithm able to effectively locate and characterize the region of interest (Figure~\ref{fig:sample_function_sampling}), but that the performance at both low and high $T$ is important in locating the global minimum.

Although simulated annealing is effective at locating the general area of the peak, since it distorts the likelihood over the course of the run (thus violating ``detailed balance''), samples cannot be used to reconstruct the associated PDF. Our solution is to only include simulated annealing during the burn-in phase to help locate the global minimum. Afterwards, we set $T(t)$ to 1, re-spawn an ensemble of walkers in an $N$-dimensional multivariate Gaussian distribution around the best-fit value located by the ensemble, and re-construct $P(z)$ from a subset of the latter trials.

As mentioned in \S\ref{subsubsec:mcmc}, in almost all cases MCMC-based methods have a difficult time effectively characterizing multimodal spaces. In particular, the efficiency of ensemble MCMC sampling rapidly declines when applied to multi-modal surfaces, especially when the modes are sharply peaked \citep{foremanmackey+13}. While respawning the ensemble of walkers around the best-fit set of parameters will by default miss truly multimodal $P(z)$'s, we have chosen this approach as a compromise to give a slightly biased view of the ``true'' $P(z)$ rather than completely undersampling and/or mischaracterizing it.

\begin{figure*}
	\includegraphics[scale=0.69]{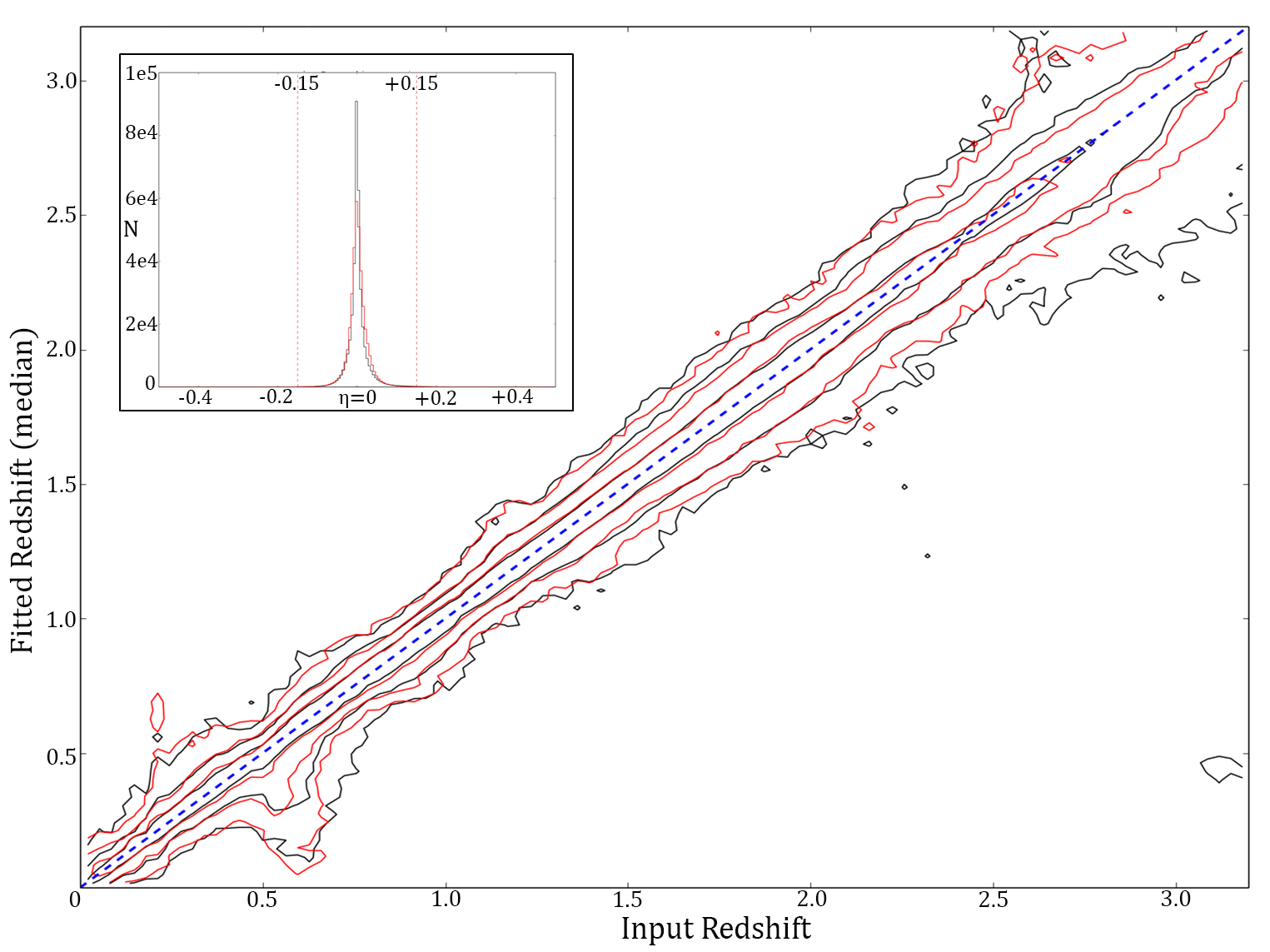}
	\caption{The distribution of input redshifts $\rom{z}{in}$ versus fitted redshifts $\rom{z}{fit}$ (70\%, 95\%, and 99\% contours plotted) for our brute-force code (black) and our MCMC-based, simulated annealing-driven \texttt{BAD-Z} code (red) using a grid of $\sim$\,2 million elements ($\Delta z = 0.005$) at $z \leq 3.2$. The marginalized distribution of the fractional redshift error $\eta \equiv | \rom{z}{fit} - \rom{z}{in} | / (1 + \rom{z}{in})$ is shown in the upper left inset with the same color scheme, with the usual threshold for ``catastrophic errors'' ($\rom{\eta}{cat}=0.15$) indicated with dashed red lines. The catastrophic outlier fractions for the two samples are 1.0\% and 0.7\%, respectively. In addition to being $\sim$\,40 times more efficient than the brute-force approach, \texttt{BAD-Z} provides similar levels of overall accuracy while retaining the ability to accurately capture the underlying redshift PDF.}
	\label{fig:badz}
\end{figure*}

We term our photo-z algorithm \textbf{B}risk \textbf{A}nnealing-\textbf{D}riven Redshifts (\textbf{Z}) (\texttt{BAD-Z}), which is summarized below:
\begin{enumerate}
	\item Initialize $W$ walkers on an arbitrary $N$-dimensional input grid drawn from a uniform distribution.
	\item Start a simulated annealing-driven ensemble MCMC run with a given set of $T(t=0)$ and $\Delta T/\Delta(\textrm{ensemble run})$ values to find the region of the global minimum.
	\item Spawn a new distribution of $W$ walkers in an $N$-dimensional multivariate Gaussian with some fractional spread $\rom{\sigma}{frac}$ in each dimension around the trial with the highest likelihood from the previous simulated annealing-driven run.
	\item Set $T(t)=1$ and $\Delta T/\Delta(\textrm{ensemble run})=0$ and start a new run for an additional $\rom{N}{MCMC}$ number of ensemble runs.
	\item After discarding an initial fraction $\rom{f}{disc}$ of trials from this second run, reconstruct the  redshift PDF using the remaining accepted trials.
\end{enumerate}

\texttt{BAD-Z} is written \texttt{C} and parallelized using \texttt{OpenMP}.

\section{Testing Performance against a Grid-based Approach}
\label{sec:badz_test}

In order to showcase the performance of \texttt{BAD-Z} relative to an appropriate baseline, we create a brute-force counterpart (also coded in \texttt{C} and parallelized using \texttt{OpenMP}) that functions exactly as described in \S\ref{subsubsec:grid}, a process that involves a total of $\sim$\,$\sn{1.6}{6}$ trials per object. We run both codes on the mock photometric catalog described in \S\ref{sec:mockcatalog} using the same grid outlined in \S\ref{sec:mapping}.


While our brute-force code by definition fits the entire grid to each individual object, \texttt{BAD-Z} traverses this grid utilizing an ensemble with $\rom{N}{walkers}=150$, $a=2$, $T(t=0)=3.0$, $\Delta T/\Delta (\textrm{ensemble run}) = 0.03$, $\rom{\sigma}{frac}=0.15$, $\rom{N}{MCMC}=200$, and $\rom{f}{disc}=0.4$. Note that this is a deliberately \textit{conservative} choice of parameters that leads to a total of 45,150 trials (2.7\% of the full grid) per object (i.e., 301 ensemble runs), of which only 18,000 ($1.1\%$ of the full grid) are used in the reconstruction of the final PDF. An example run for an individual object in our mock catalog is shown in Figure~\ref{fig:badz_run}.

Using the reconstructed $P(z)$ for each individual object, we classify the best-fitting redshift as the median of the distribution $\rom{z}{med}=\textrm{median}(P(z))$. The 2-D distributions of input redshift $\rom{z}{in}$ versus fitted redshift $\rom{z}{fit}=\rom{z}{med}$ for both methods are shown in Figure~\ref{fig:badz}. Although \texttt{BAD-Z} samples $\sim$\,40 times less than its brute-force counterpart, it produces photo-z estimates with comparable accuracy to the brute-force approach and similar catastrophic outlier ($| \rom{z}{fit} - \rom{z}{in} | / (1 + \rom{z}{in}) > 0.15$) rates (0.7\% and 1.0\% for \texttt{BAD-Z} and brute-force, respectively).\footnote{We find similar results using redshift estimates derived from the mean and mode of each $P(z)$, which are not shown.}

In addition, we find that \texttt{BAD-Z} is quite effective at finding good fits to the data, locating ``optimal'' ($\chi^2_{\textrm{fit}}/\chi^2_{\textrm{base}} \leq 1$) fits in $\sim$\,$55\%$ of cases and ``reasonable'' ($\chi^2_{\textrm{fit}}/\chi^2_{\textrm{base}} \leq 5$) ones in $\sim$\,$90\%$. As the resulting redshifts are accurate in the majority of fitted objects (Figure~\ref{fig:badz}), this indicates that even in cases where \texttt{BAD-Z} fails to find the ``best'' fit to the data (for a variety of possible reasons), it still manages to probe the surrounding region to high enough accuracy that the marginalized $P(z)$ distribution gives accurate predictions.

Finally, we compare the similarity between the $P(z)$'s computed from \texttt{BAD-Z} to those computed with our brute-fore code to determine the general ability of our new code to recover accurate PDFs as well as the overall redshift distribution. The [16th, 50th, 84th] quartiles of the difference between the median $P(z)$ solutions computed with both codes is [-3\%, -0.5\%, 1.5\%], indicating that while \texttt{BAD-Z} has a slight negative bias, the overall agreement between both codes is excellent. We do find however, that the distribution about the correct redshifts is slightly broader for \texttt{BAD-Z} than for our brute-force code, with a two-sample Kolmogorov-Smirnov (KS) test statistic of $p=0.03$ (2.2-$\sigma$).

To confirm this, we investigate the distribution of errors for individual objects, finding that the PDFs computed by \texttt{BAD-Z} are indeed preferentially broadened relative to the brute-force code by $\sim$\,50\% (median). This is likely caused by the limited number of samples used to reconstruct the PDF (due to our high choice for $\rom{f}{disc}$) and the discrete nature of the grid (which can lead to a ``pile-up'' of our walkers at specific grid points, reducing our effective resolution). Ultimately, we find both codes produce $N(z)$ distributions that are consistent with being drawn from the same parent population as the spec-z distribution and are consistent with being drawn from the same parent population relative to each other (KS $p$-values $> 0.05$ in all cases), indicating that although the PDFs provided by \texttt{BAD-Z} might be ``rougher'' than those derived from sampling the entire grid they still provide reliable measurements.

In summary, by understanding the general topology of the relevant photo-z likelihood surface seen by a specific grid of pre-computed model photometry, we are able to design an algorithm that is substantially more efficient than traditional brute-force approaches while retaining similar levels of accuracy. Our ensemble MCMC-based, simulated annealing-driven \texttt{BAD-Z} algorithm performs extremely well on the bumpy, degenerate photo-z likelihood surface explored in this work, giving good fits in $>$\,$90\%$ of cases with a comparable catastrophic outlier fraction ($0.7\%$ versus $1.0\%$). \texttt{BAD-Z} thus illustrates the power inherent in the general methodology outlined in this section.\footnote{The code also runs relatively quickly, providing redshift PDFs to individual objects in $\sim 1.5$\,s per core and to the entire mock catalog (parallelized with 32 threads) in a little over five hours. Significant gains could likely be achieved with optimization of the relevant code.}

Although these results are promising, we wish to emphasize that \texttt{BAD-Z} is subject to the limitations inherent in all MCMC sampling approaches regarding the stochastic reconstruction of PDFs as a function of sample size and the limitations of locating and converging to the target distribution during burn-in. While we have attempted to bypass these problems through the use of ensemble MCMC sampling (to allow for more exploration of the parameter space and better reconstruction of the PDF), simulated annealing (to help ensure robust convergence to the global minimum), and the large width of our ``respawned'' distribution after burn-in (with a 1-$\sigma$ width of 15\% of the corresponding parameter space), our chosen method by construction \textit{will} mis-characterize truly multi-modal distributions, especially those with a large number of widely-separated modes with similar amplitudes.

\section{Conclusion}
\label{sec:conclusion}

While photometric redshifts (photo-z's) represent an integral part of modern extragalactic science, outstanding issues that currently plague template fitting-based approaches are concerning in the face of looming future ``big data''-oriented surveys. This work represents the first steps towards moving template-fitting photo-z codes from a runtime-limited regime to a memory-limited one. Our main results are as follows:
\begin{enumerate}
	\item Using a pre-generated grid of $\sim$\,2 million elements ($\Delta z=0.005$), we create ``maps'' of the associated photo-z likelihood surfaces. For our chosen grid, we find that the surface is significantly ``bumpy'', with a substantial number of minima occupying large areas compared to the region directly surrounding the global best-fit value.
	\item We explore the use of ``greedy'' minimization algorithms on 2-D redshift-reddening slices to supplement photo-z searches. We find these 2-D slices remain significantly multi-modal such that locating the best-fit model proves to be difficult even with the help of metaheuristics such as basin-hopping.
	\item Building on these results, we design a specific algorithm \textbf{B}risk \textbf{A}nnealing-\textbf{D}riven Redshits (\textbf{Z}) (\texttt{BAD-Z}) to explore pre-generated grids of model photometry during photo-z searches through a combination of ensemble MCMC sampling and simulated annealing.
	\item Using a mock catalog of 384,662 COSMOS galaxies, we test the performance of \texttt{BAD-Z} over a wide wavelength (UV$ugrizYJH$) and redshift ($0 < z \lesssim 3.2$) range. Compared to a grid-based counterpart, we find \texttt{BAD-Z} is $\sim$\,40 times more computationally efficient, retains a similar level of accuracy, and performs robustly over the entire redshift range probed.
\end{enumerate}

These results are merely the first steps towards a rigorous attempt to improve photo-z's and only part of a much larger, extended effort within the extragalactic astronomical community. For instance, almost all photo-z codes -- including the one showcased in this work -- utilize exclusively color information to derive $P(z)$. This, however, ignores potentially important information such as clustering, morphology, angular size, and/or surface brightness, all of which might either improve accuracy or help distinguish the dominant mode(s) of a multimodal redshift PDF. In particular, incorporating clustering information would be a useful next step towards exhausting all available information contained in photometric surveys \citep{menard+13,newman+15}.

In addition, this work has focused almost exclusively on template-fitting approaches to deriving photo-z's, bypassing  a wide range of machine learning techniques that are almost certainly critical in improving upon current photo-z methodologies. In particular, (un)supervised machine learning approaches such as Self-Organizing Maps (\citealt{kohonen82,kohonen01,carrascokindbrunner14}; Masters et al. 2015, submitted) offer the opportunity to move beyond simple inverse mapping approaches to instead incorporate prior knowledge about a given dataset in increasingly sophisticated ways \citep{dahlen+13,carrascokindbrunner14b}.

Finally, while this work has focused on many of the more computationally-oriented avenues towards improving photometric redshifts, a major unresolved issue in current photo-z searches is the dual set of model uncertainties that arise due to the use of local galaxy templates (and emission line scaling relations) to probe galaxies at much higher redshifts and the limited range of dust templates (and lack of priors; \citealt{repp+15}) often used during the fitting process. Both of these avenues must be explored further in order to develop an improved set of templates and a better understanding of modeling uncertainties. Ultimately, there remain ample opportunities for further investigation in advancing new techniques, creating superior template sets, and improving overall computational efficiency.

\section*{Acknowledgements}

JSS would like to thank Douglas Finkbeiner and Zachary Slepian for comments that improved the quality of this work and Charles Alcock for supervising the senior thesis course where much of this work was completed. JSS is grateful for financial support from the Herchel Smith-Harvard Summer Undergraduate Research Fellowship, the Harvard University Department of Astronomy, the Harvard College Observatory, and CREST funding from the Japan Science and Technology Agency (JST). This work has benefited extensively from access to computing resources at IPAC and Harvard.



\bsp	
\label{lastpage}
\end{document}